\documentclass[12pt]{article}
\usepackage{latexsym}
\begin{document}

\title{{\bf EXTENDED\\ ELECTRODYNAMICS}:\\
I. Basic Notions, Principles and Equations}
\author{{\bf S.Donev}\\
Institute for Nuclear Research and Nuclear Energy,\\
Bulg.Acad.Sci., 1784 Sofia, blvd.Tzarigradsko shausee 72\\
e-mail: sdonev@inrne.acad.bg\\ BULGARIA \and \\
{\bf M.Tashkova}\\
Institute of Organic Chemistry with Center of \\ Phytochemistry,
Bulg.Acad.Sci., 1574 Sofia,\\
Acad. G.Bonchev Str., Bl. 9\\BULGARIA}

\date{}

\maketitle

\begin{abstract}
This paper aims to present an elaborate view on the motivation and
realization of the idea to extend Maxwell's electrodynamics to \linebreak
{\it Extended Electrodynamics} in a reasonable and appropriate way
in order to make it possible to describe
electromagnetic
(3+1)-soliton-like objects in vacuum and in the presence of continuous media
(external fields) [15]-[20], exchanging energy-momentum with the
electromagnetic field.

\end{abstract}

\newpage

\noindent {\bf \large 1. Preliminary notes}
\vskip 0.5cm
Classical Electrodynamics (CED) [1] is probably the most fascinating and
complete part of the Classical Field Theory. Intuition, free thought,
perspicuity and research skill of many years finally brought about the
synthesis of experiment and theory, of physics and mathematics, which we
have been calling for short the {\it Maxwell equations} for the duration
of a century and a half. From the beginning of the second half of the 19th
century till its end these equations turned from abstract theory into
daily practice, as they are today. Their profound study during the first
half of the 20th century brought forward a new theoretical concept in
physics known as {\it relativism}. Brave and unprejudiced scientists
enriched and widened the synthesis achieved through the Maxwell equations,
and created a new synthesis known briefly as {\it quantum
electrodynamics}.  Every significant scientific breakthrough is based on
two things: {\it respect for the workers and their work and respect for
the truth}. "May everyone be respected as a personality, and nobody as an
idol" one of the old workers used to say.  We may paraphrase that saying:
"may every scientific truth be respected, but no one be turned into
dogma".

In our attempt to extend CED we tried to follow the values this creed
teaches, as far as our humble abilities allow us to.  Together with the
analysis of the classical electrodynamics and the discrete conception for
the structure of the electromagnetic field, the path followed brought us
to the conclusion that a {\it soliton-like solution of appropriate
non-linear equations characterized by an intrinsic periodical process is
the most adequate mathematical model of the basic structural unit of the
field - the photon}. The fact that neither the Maxwell equations nor the
quantum electrodynamics offer the appropriate tools to find such
solutions, unmistakably emphasizes the necessity to look for new
equations.  We decided not to follow the usual way for nonlinearization of
CED [2], because, from our point of view, it does not comprise
sufficiently fruitful and new ideas. Other attempts in this direction one
may find in [3]-[12].

 The leading physical ideas  in our approach were the dual
("electro-\linebreak magnetic") nature of the field on the one hand and
the local Energy-\linebreak Momentum Conservation Laws on the other. The
realization that every such soliton-like solution determines in an
invariant way its own {\it scale factor}, as well as the suitable
interpretation of the famous formula of Planck for the relation between
the full energy $E$ of the photon and the frequency $\nu=1/T$ of the
beforementioned periodical process, which we prefer to write down as
$h=E.T$, helped us to formulate the rules for separating the realistic
soliton-like models of the photon from among the rest. The resulting
soliton-like solutions [19] possess all integral qualities of the photon,
as described by quantum electrodynamics, but also a structure, organically
tied to an intrinsic periodical process, which in its turn generates an
intrinsic mechanical angular momentum - spin (helicity). We consider this
soliton- like oscillating non-linear wave much more clear and
understandable than the {\it ambiguous "particle-wave" duality}.

The dual 2-component nature of the field predetermined to a great extent
the generalization of the equations in the case of an interaction with
another continuous physical object, briefly called medium [20].  The
proposed physical interpretation of the classical Frobenius equations for
complete integrability [13],[14] of a system of non-linear partial
differential equations as a {\it criterion for the absence of
dissipation}, turned out to be relevant and was effectively used. The
fruitfulness of the new non-linear equations is clearly shown by the
family of solutions, giving (3+1)-dimensional interpretation of all
(1+1)-dimensional 1-soliton and multisoliton solutions [20]. We note that
making use of differential geometry proved extremely useful.

This paper follows the main track of our efforts to build a clear and
consecutive picture of motivations and theoretical results.
While most of the solutions to the new nonlinear equations we have found
have been already published
[15]-[20], a consecutive and well motivated reasoning, leading to the new
equations is still missing. So,
 the stress in this paper is laid on the conceptual and generic framework.
The purpose is to bring the reader to the conviction that the extension of
CED, developed by us during the last 10 years, is necessary and is
physically and mathematically quite natural. We do not present solutions to
equations.  We set the problem to build a description of electromagnetic
soliton-like objects, then we consider step by step physical arguments and
try to find the corresponding step by step mathematical adequacies. Our
belief is that if the steps are in the right direction the positive results
come inevitably. In its turn every new positive result (solution or
relation), suggests new insights and invites to search for new more definite
quantities and relations. This, step by step creative process, brought us to
the today's motivational look on what we call {\it Extended Electrodynamics}
(EED). It worths from time to time to reconsider and re-estimate the
importance and significance of any reason and idea been used, because this
gives birth to new reasons and ideas and helps to sift out the basic and
meaningful from the occasional and nonsignificant.

\vskip 1cm
\noindent {\bf \large 2. Physical conception for the electromagnetic field \\
in Extended Electrodynamics}
\vskip 0.5cm
As it is well known, the mathematical models of the real vacuum
electromagnetic fields in Classical Electrodynamics  are
"infinite", or if they are finite, they are strongly time-unstable. These
models are not consistent with a number of experimental facts. A deeper
analysis of CED, carried out in the first third of this century, resulted in
the new conception for a discrete character of the field. This physical
understanding of the field is the true foundation of EED and it shows clearly
the principal differences between CED and EED.  For the sake of clarity we
shall formulate our point of view more explicitly.

{\it The electromagnetic field in vacuum is of discrete character and
consists of single, not-interacting (or very weakly interacting) finite
objects, called photons. All photons move uniformly as a whole by the same
velocity 'c', carry finite energy '$E$', momentum '{\bf p}' and intrinsic
angular momentum. These features imply structure and internal periodic
process of period '$T$', which may be different for the various photons. The
quantity '$E.T$', called "elementary action", has the same value for all
photons and is numerically equal to the Planck constant '$h$'. The invariance
of '$c$' and '$h$' means nondistinguishability of the photons, considered as
invariant objects.  The integral value of the intrinsic angular momentum is
equal to '$\pm h$'. For the topology of the 3-dimensional region, occupied by
the photon at any moment, there are no experimental data, so it is desirable
that the model-solutions to allow arbitrary initial data}.

We'd like to stress the following: EED considers photons as {\it real
finite objects}, and {\it not as convenient theoretical concepts}, and it
aims to build adequate mathematical models of these entities. So, the
first important problem is to point out the algebraic character of the
mathematical object describing a {\it single} photon. The corresponding
generalization for a number of photons is then easily done.

\vskip 1cm
\noindent{\bf \large 3. Choice of the modelling mathematical object}
\vskip 0.5cm
According to the non-relativistic formulation of CED the electromagnetic
field has two aspects: "electric" and "magnetic". These two aspects of the
field are described by two 1-forms on ${\cal R}^3$ and a  parametric
dependence on time is possible: the electric field $E$ and the magnetic field
$B$. The following considerations will bring us to the conclusion that
these two fields can be considered as two {\it vector components} of a new
object, 1-form $\Omega$, {\it taking values in a real 2-dimensional vector
space}, naturally identified with ${\cal R}^2$. In fact, let's consider
the question: do there exist constants $(a,b,m,n)$, such that the linear
combinations
\[
E'=aE+bB,\ B'=mE+nB
\]
give again a solution? The answer to this question is positive
iff $m=-b, \linebreak
n=a$. The new solution will have energy density $w'$
and momentum ${\bf S'}$ as follows:
\[
w'=\frac {1}{8\pi}\biggl((E')^2 +
(B')^2\biggr)= \frac {1}{8\pi}(a^2 + b^2)\biggl(E^2 + B^2\biggr),
\]
\[{\bf S'}=(a^2+b^2)\frac {c}{4\pi} E\times B.
\]
Obviously, the new and the old
solutions will have the same energy and momentum if $a^2+b^2 =1$.

This simple but important observation shows that besides the usual linearity,
Maxwell's equations admit also "cross"-linearity, i.e. linear combinations of
$E$ and $B$ of a definite kind define new solutions. Therefore, the
difference between the electric and magnetic fields becomes non-essential.
The important point is that with every solution $(E,B)$ of Maxwell's
equations a 2-dimensional real vector space, spanned by the couple $(E,B)$,
is associated, and the linear transformations, which transform solutions into
solutions, are given by matrices of the kind
\[
\left\|\matrix{a   &b\cr
	       -b  &a\cr} \right\|.
\]
If these matrices are unimodular, i.e. if $a^2+b^2=1$, the energy and
momentum do not change. It is well known
that matrices of this kind do not change the canonical complex structure
$J$ in ${\cal R}^2$.(Recall that  if the canonical basis of ${\cal R}^2$ is
denoted by $(e_1,e_2)$  then $J$ is defined by $J(e_1)=e_2$, $J(e_2)=-e_1$.)

The above remarks suggest to consider $E$ and $B$ as two vector-components
of an ${\cal R}^2$-valued 1-form $\Omega$:
\[
\Omega=E\otimes e_1 + B\otimes e_2.
\]
So, the current $ {\bf j}$ becomes 1-form ${\cal J}={\bf j}\otimes e_1$ with
values in ${\cal R}^2$, and the charge density becomes a function
${\cal Q}=\rho\otimes e_1$ with values in ${\cal R}^2$. Maxwell's equations
\begin{equation}
\frac 1c \frac{\partial E}{\partial t}=rotB - \frac {4\pi}{c}{\bf j},
\quad divB=0, 
\end{equation}
\begin{equation}
\frac 1c \frac{\partial B}{\partial t}=-rotE,\quad \ \ divE=4\pi\rho. 
\end{equation}
take the form
\begin{equation}
\frac {1}{c} \frac {\partial \Omega}{\partial t}=
-\frac {4\pi}{c}{\cal J}-*{\bf d}J(\Omega),\quad
\delta \Omega=4\pi {\cal Q}, 
\end{equation}
\noindent where
$J(\Omega)=E\otimes J(e_1)+B\otimes J(e_2)=E\otimes e_2-B\otimes e_1$
and {\bf d} is the exteriour derivative.
Note that according to the sense of the concept of current in CED and
because of the lack of magnetic charges, it is necessary to exist a basis
in ${\cal R}^2$, in which ${\cal J}$ and ${\cal Q}$ to have components
only along $e_1$. Nevertheless, this point of view shows that even at this
non-relativistic level the separation of the $EM$-field to "electric" and
"magnetic" is not adequate to the real situation.  The mathematical object
$\Omega$ unifies and, at the same time, distinguishes the two sides of the
field: there is a basis in ${\cal R}^2$, in which the electric and
magnetic components are delimited, but in an arbitrary basis the two
components mix (superimpose), so the difference between them is deleted.

In the relativistic formulation of CED the difference between the electric
and magnetic components of the field is already quite conditional, and from
invariant-theoretical point of view there is no difference. However, the
2-component character of the field acquires a new meaning and manifests
itself differently. In order to come to this we make the following
considerations.

As we mentioned above, some linear
combinations of the electric and magnetic fields generate a new solution to
Maxwell's equations. In particular, the transformation, defined by
the matrix
\[
\left\|\matrix{0   &1\cr
	       -1  &0\cr} \right\|,
\]
defining a complex structure in ${\cal R}^2$, transforms a solution
of the kind $(E,0)$ into a new solution of the kind $(0,E)$ and a solution
of the kind $(0,B)$ into a solution of the kind $(-B,0)$, i.e. the electric
component into magnetic and vice versa.
This observation draws our attention to looking
for a complex structure $J,J^2=-id$ in the bundle of 2-forms on the Minkowski
space $M$, such that if $F$ presents the first component of the field, then
$J(F)$ to present the second component of the same field. Such complex
structure really exists, in fact, it coincides with the
restriction of the Hodge $*$-operator, defined by the pseudometric
$\eta$ through the equation
$$
\alpha\wedge *\beta=-\eta (\alpha,\beta)\sqrt{|\eta|}dx\wedge dy\wedge
dz\wedge d\xi =-\eta(\alpha,\beta)\omega_o,\ \xi=ct,
$$
to the space of 2-forms: $**_2=-id_{\Lambda^2(M)}$. So, the
non-relativistic vector components $(E,B)$ are replaced by the relativistic
vector components $(F,*F)$. The following considerations support also such a
choice.

The relativistic Maxwell's equations in vacuum ${\bf d}F=0,\ {\bf d}*F=0$
are, obviously, invariant with respect to the interchange $F\rightarrow *F$.
Moreover, if $F$ is a solution, then an arbitrary linear combination $aF+b*F$
is again a solution. More generally, if $(F,*F)$ defines a solution, then the
transformation $(F,*F)\rightarrow (aF+b*F,mF+n*F)$ defines a new solution for
an arbitrary matrix
\[
\left\|\matrix{a   &m\cr
	       b   &n\cr} \right\|.
\]
Now, using the old notation $\Omega$ for the new object
$\Omega=F\otimes e_1+*F\otimes e_2$, Maxwell's equations are written down
as ${\bf d}\Omega=0$, or equivalently $\delta \Omega=0$, where the
coderivative operator $\delta=*{\bf d}*$ is just the (minus) {\it
divergence}: $(\delta F)_\mu=-\nabla_\nu F^\nu_\mu $.
Clearly, an arbitrary linear
transformation of the basis $(e_1,e_2)$ keeps $\Omega$ as a solution.

Recall now the energy-momentum tensor in CED, defined by
\begin{equation}
Q_\mu^\nu=\frac {1}{4\pi}\biggl[\frac 14
F_{\alpha\beta}F^{\alpha\beta}\delta_\mu^\nu -
F_{\mu\sigma}F^{\nu\sigma}\biggr]=\\
\frac {1}{8\pi}\biggl[-F_{\mu\sigma}F^{\nu\sigma}-
(*F)_{\mu\sigma}(*F)^{\nu\sigma}\biggr].
\end{equation}
It is quite clearly seen, that $F$ and $*F$ participate in the same way in
$Q_\mu^\nu$, and the full energy-momentum densities of the field are obtained
through summing up the energy-momentum densities, carried by $F$ and $*F$.
Since the two expressions $F_{\mu\sigma}F^{\nu\sigma}$ and
$(*F)_{\mu\sigma}(*F)^{\nu\sigma}$ are not always equal, the distribution
of energy-momentum between $F$ and $*F$ may change in time, i.e.
energy-momentum may be transferred from $F$ to $*F$, and vice versa. So we
may interpret this phenomenon as a special kind of interaction between $F$
and $*F$, responsible for some internal redistribution of the field
energy. Now, in vacuum it seems naturally to expect, that the
energy-momentum, carried from $F$ to $*F$ in a given 4-volume, is the same
as that, carried from $*F$ to $F$ in the same volume. However, in presence
of an active external field, exchanging energy-momentum with $\Omega$, it
is hardly reasonable to trust the same expectation just because of the
specific structure the external field (medium) may have. So, the external
field (further  any such external field will be called also {\it medium}
for short) may exchange energy-momentum preferably by $F$ or $*F$, as well
as it may support the internal redistribution of the field energy-momentum
between $F$ and $*F$, favouring $F$ or $*F$. From the explicit form of the
energy-momentum tensor it is seen that the field may participate in this
exchange by means of any of the two terms $F_{\mu\sigma}F^{\nu\sigma}$ and
$(*F)_{\mu\sigma}(*F)^{\nu\sigma}$.  Moreover, for the divergence of the
energy-momentum tensor we easily obtain
\begin{equation}
\nabla_\nu Q_\mu^\nu=\frac {1}{4\pi}\biggl[F_{\mu\nu}(\delta
F)^\nu+(*F)_{\mu\nu}(\delta *F)^\nu\biggr].
\end{equation}
It is clearly
seen that the quantities of energy-momentum, which any of the two components
$F$ and $*F$ may exchange in a unit 4-volume are {\it invariantly} separated
and  given respectively by
$$
F_{\mu\nu}(\delta F)^\nu, \quad (*F)_{\mu\nu}(\delta *F)^\nu.
$$
But, in CED the exchange through $*F$ {\it is forbidden}, the expression
\linebreak
$(*F)_{\mu\nu}(\delta *F)^\nu$ is always, in {\it all media}, equal to
zero.  This comes from the unconditional assumption, that the Faraday's
induction law is universally true.  Of course, we do not reject the existence
of media, not allowing energy-momentum exchange through $*F$,
but we do not share the opinion that all media behave in this
same way.  On the other hand, in case of vacuum, we can not delimit $F$ from
$*F$, these are two solutions of the same equation and it is all the same
which one will be denoted by $F$ (or $*F$), i.e. CED does not give an
intrinsic criterion for a respective choice. Only in regions with non-zero
free charges and currents, when ${\bf d}F=0$ and $\delta F=4\pi j\neq 0$, the
choice can be made, but this presupposes (postulates) that the field is able
to interact, i.e. to exchange energy-momentum, {\it only} with charged
particles, i.e. through $F$. This postulate we can not assume ad hoc.

Having in view these considerations we assume the following postulate in EED
in order to specify the algebraic character of the modelling mathematical
object:

\vskip 0.5cm
{\it In EED the electromagnetic field is described by a 2-form $\Omega$,
defined on the Minkowski space-time and valued in a real 2-dimensional vector
space ${\cal W}$ and such, that there is a basis $(e_1,e_2)$ of ${\cal W}$ in
which $\Omega$ takes the form}
\begin{equation}
\Omega = F\otimes e_1 + *F\otimes e_2.
\end{equation}

\vskip 0.5cm
Since ${\cal W}$ is isomorphic to ${\cal R}^2$, further we shall write only
${\cal R}^2$ and all relations obtained can be carried over to ${\cal W}$ by
means of the corresponding isomorphism. In particular, every ${\cal W}$ will
be considered as being provided with a complex structure $J$, so, the
group of automorphisms of $J$ is defined. Our purpose now is to prove that
the set of 2-forms of the kind (6) is stable under the invariance group of
$J$.

First we note, that the equation $aF+b*F=0$ requires $a=b=0$. In fact, if
$a\neq 0$ then $F=-\frac{b}{a}*F$. From $aF+b*F=0$ we get $a*F-bF=0$ and
substituting $F$, we obtain $(a^2+b^2)*F=0$, which is possible only if
$a=b=0$ since $*F\neq 0$. In other words, $F$ and $*F$ are linearly
independent.
Let now $(k_1,k_2)$ be another basis of ${\cal R}^2$ and consider the
2-form $\Psi =G\otimes k_1+*G\otimes k_2$. We express $(k_1,k_2)$ through
$(e_1,e_2)$ and take in view what we want:
\[
G\otimes k_1+*G\otimes k_2=G\otimes (ae_1+be_2)+*G\otimes (me_1+ne_2)=
\]
\[
=(aG+m*G)\otimes e_1+(bG+n*G)\otimes e_2=(aG+m*G)\otimes e_1+
*(aG+m*G)\otimes e_2.
\]
Consequently, $bG+n*G=a*G-mG$, i.e. $(b+m)G+(n-a)*G=0$, which
requires $m=-b,\ n=a$, i.e. the transformation matrix is
\[
\left\|\matrix{a   &-b\cr
	       b   &a\cr} \right\|.
\]
Besides, if ${\Omega}_1$ and ${\Omega}_2$ are of the kind (6), it is
easily shown that the linear combination $\lambda\Omega_1+\mu\Omega_2$ is of
the same kind (6). These results show that the 2-forms of the kind (6)
form a stable with respect to the automorphisms of $({\cal R}^2,J)$ subspace
of the space $\Lambda ^2(M,{\cal R}^2)$. Moreover, if the component $F$ is
chosen, then the basis $(e_1,e_2)$ is uniquely determined. In other words,
every $\Omega$ of the kind (6) allows various representations of this kind,
but every component $F$ determines unique basis
$\left(e_1(F),e_2(F)\right)$, and so 2 subspaces $\{e_1\}$ and $\{e_2\}$ of
${\cal R}^2$. (Further the argument ($F$) of the corresponding basis is
omitted).

In order to separate the class of bases we are going to use,
first we recall the product of 2 vector valued differential forms. If
$\Phi$ and $\Psi$ are respectively $p$ and $q$ forms on the same manifold $N$,
taking values in the vector spaces $W_1$ and $W_2$ with corresponding bases
$(e_1,...,e_m)$ and $(k_1,...,k_n)$, and
$\varphi :W_1\times W_2\rightarrow W_3$ is a bilinear map into the vector
space $W_3$, then a $(p+q)$-form $\varphi\left(\Phi,\Psi\right)$ on $N$ with
values in $W_3$ is defined by
\[
\varphi\left(\Phi,\Psi \right)=
\sum_{i,j}\Phi^i\wedge \Psi^j \otimes \varphi(e_i,k_j).
\]
In particular, if $W_1=W_2$ and $W_3={\cal R}$, and the bilinear map is
scalar (inner) product $g$, we get
\[
\varphi\left(\Phi,\Psi \right)=\sum_{i,j}\Phi^i\wedge \Psi^j g_{ij}.
\]

Let now $X$ and $Y$ be 2 arbitrary vector fields on the Minkowski space $M$,
$\Omega$ be of the kind (6), $Q_{\mu\nu}$ be the energy tensor in CED and
$g$ be the canonical inner product in ${\cal R}^2$. Then the class of bases
we shall use will be separated by the following equation \begin{equation}
Q_{\mu\nu}X^\mu Y^\nu=\frac12 *g\Bigl(i(X)\Omega,*i(Y)\Omega\Bigr).
\end{equation}
We develop the right hand side of this equation and obtain
\[
\frac12 *g\Bigl(i(X)\Omega,*i(Y)\Omega\Bigr)=
\]
\[
\frac12 *g\Bigl(i(X)F\otimes e_1+i(X)*F\otimes e_2,*i(Y)F\otimes e_1
+ *i(Y)*F\otimes e_2\Bigr)=
\]
\[
=\frac12*\biggl[\Bigl(i(X)F\wedge*i(Y)F\Bigr)g(e_1,e_1)+
\Bigl(i(X)F\wedge*i(Y)*F\Bigr)g(e_1,e_2)+
\]
\[
+\Bigl(i(X)*F\wedge*i(Y)F\Bigr)g(e_2,e_1)+
\Bigl(i(X)*F\wedge*i(Y)*F\Bigr)g(e_2,e_2)\biggr]=
\]
\[
=-\frac12 X^\mu Y^\nu\biggl[F_{\mu\sigma}F_\nu^\sigma g(e_1,e_1)+
(*F)_{\mu\sigma}(*F)_\nu^\sigma g(e_2,e_2)+
\]
\[
+\Bigl(F_{\mu\sigma}(*F)_\nu^\sigma+(*F)_{\mu\sigma}F_\nu^\sigma\Bigr)
g(e_1,e_2)\biggr]=
-\frac12 X^\mu Y^\nu\biggl[F_{\mu\sigma}F_\nu^\sigma +
(*F)_{\mu\sigma}(*F)_\nu^\sigma\biggr].
\]
In order this relation to hold it is necessary to have
\[
g(e_1,e_1)=1,\ g(e_2,e_2)=1,\ g(e_1,e_2)=0,
\]
i.e., we are going to use {\it orthonormal} bases. So, the stability group of
the subspace of forms of the kind (6) is reduced to $SO(2)$ or $U(1)$. So,
in this approach, the group $SO(2)$ appears in a pure algebraic way, while in
the gauge interpretation of CED this group is associated with the equation
\linebreak ${\bf d}F=0$, i.e. with the traditional and not shared by us
understanding, that the $EM$-field can not exchange energy-momentum with any
medium \linebreak through $*F$.

\vskip 1cm
\noindent{\bf \large 4. Differential equations for the field}
\vskip 0.5cm
We proceed to the main purpose, namely, to write down differential equations
for our object $\Omega$, which was chosen to model the $EM$-field. We shall
work in the orthonormal basis $(e_1,e_2)$, where the field has the form
(6). The two vectors of this basis define two mutually orthogonal
subspaces $\{e_1\}$ and $\{e_2\}$, such that the space ${\cal R}^2$ is a
direct sum of these two subspaces: \ ${\cal R}^2=\{e_1\}\oplus\{e_2\}$. So,
we have the two projection operators $\pi_1:{\cal R}^2\rightarrow \{e_1\},\
\pi_2:{\cal R}^2\rightarrow \{e_2\}$. These two projection operators extend
to projections in the ${\cal R}^2$-valued differential forms on $M$:
\[
\pi_1 \Omega =\pi_1(\Omega^1\otimes k_1+\Omega^2 \otimes k_2)=
\Omega^1 \otimes \pi_1 k_1 +\Omega^2 \otimes \pi_1 k_2=
\]
\[
=\Omega^1 \otimes \pi_1(ae_1+be_2) + \Omega^2\otimes \pi_1(me_1+ne_2)=
(a\Omega^1 +m\Omega^2)\otimes e_1.
\]
Similarly,
\[
\pi_2\Omega=(b\Omega_1+n\Omega_2)\otimes e_2.
\]
In particular, if $\Omega$ is of the form (6), then
\[
\pi_1(F\otimes e_1+*F\otimes e_2)=F\otimes e_1,\
\pi_2(F\otimes e_1+*F\otimes e_2)=*F\otimes e_2.
\]

Let now our $EM$-field $\Omega$ propagates in a region, where some other
continuous physical object (external field, medium) also propagates and
exchanges energy-momentum with $\Omega$. We are going to define explicitly
the local law this exchange obeys.

First we note, that the external field is described by some mathematical
object(s). From this mathematical object, following definite rules,
reflecting the specific situation under consideration, a new mathematical
object  ${\cal A}_i$ is constructed and this new mathematical object
participates directly in the exchange defining expression. The $EM$-field
participates in this exchange defining expression directly through $\Omega$,
and since $\Omega$ takes values in ${\cal R}^2$, then ${\cal A}_i$ must also
take values in ${\cal R}^2$.

We make now two preliminary remarks. First, all operators, acting on the
usual differential forms, are naturally extended to act on vector valued
differential forms according to the rule $D\rightarrow D\times id$. In
particular,
\[
*\Omega=*(\sum_i \Omega^i\otimes e_i) = \sum_i (*\Omega^i)\otimes e_i,\
{\bf d}\Omega={\bf d}(\sum_i \Omega^i\otimes e_i) =
\sum_i ({\bf d}\Omega^i)\otimes e_i,
\]
\[
\delta \Omega=\delta(\sum_i \Omega^i\otimes e_i) =
\sum_i (\delta\Omega^i)\otimes e_i.
\]
Second, in view of the importance of the expression (5) for the divergence
of the CED energy-momentum tensor, we give its explicit deduction. Recall the
following algebraic relations on the Minkowski space:
\begin{equation}
\delta_p=(-1)^p *^{-1} {\bf d} * =*{\bf d}*,\ \delta *_p=*{\bf d}_p\ for\
p=2k+1,\ \delta *_p=-*{\bf d}_p\ for\  p=2k.
\end{equation}
If $\alpha$ is a 1-form and $F$ is a 2-form, the following relations hold:
\begin{equation}
*(\alpha\wedge *F)=-
\alpha^\mu F_{\mu\nu}dx^\nu=*\left[(*F)\wedge *(*\alpha)\right]=
\frac12 (*F)^{\mu\nu}(*\alpha)_{\mu\nu\sigma}dx^\sigma.
\end{equation}
In particular,
\[
*(F\wedge *{\bf d}F)=\frac12 F^{\mu\nu}({\bf d}F)_{\mu\nu\sigma}dx^{\sigma}=
*[\delta *F\wedge *(*F)]=(*F)_{\mu\nu}(\delta *F)^{\nu}dx^{\mu}.
\]
Having in view these relations, we obtain consecutively:
\[
\nabla_\nu Q_\mu^\nu=
\nabla_\nu\biggl[\frac 14 F_{\alpha\beta}F^{\alpha\beta}\delta_\mu^\nu-
F_{\mu\sigma}F^{\nu\sigma}\biggr]=
\]
\[
=\frac12 F^{\alpha\beta}\nabla_{\nu}F_{\alpha\beta}\delta^{\nu}_{\mu}-
(\nabla_\nu F_{\mu\sigma})F^{\nu\sigma} -
F_{\mu\sigma}\nabla_{\nu}F^{\nu\sigma} =
\]
\[
=\frac12 F^{\alpha\beta}\bigl[({\bf d}F)_{\alpha\beta\mu}-
\nabla_\alpha F_{\beta\mu}-
\nabla_\beta F_{\mu\alpha}\bigr]-
(\nabla_\nu F_{\mu\sigma})F^{\nu\sigma} - F_{\mu\sigma}\nabla_\nu
F^{\nu\sigma} =
\]
\[
=\frac12 F^{\alpha\beta}({\bf d}F)_{\alpha\beta\mu}-
F_{\mu\sigma}\nabla_\nu F^{\nu\sigma}=
-(*F)_{\mu\nu}\nabla_\sigma (*F)^{\sigma\nu}-
F_{\mu\nu}\nabla_\sigma F^{\sigma\nu}=
\]
\[
=(*F)_{\mu\nu}(\delta *F)^{\nu}+F_{\mu\nu}(\delta F)^{\nu}.
\]

Let now our field $\Omega$ interact with some other field. This interaction,
i.e. energy-momentum exchange, is performed along 3 "channels". The first 2
channels are defined by the 2 (equal in rights) components $F$ and $*F$ of
$\Omega$. This exchange is {\it real} in the sense, that some part of the
$EM$-energy-momentum may be transformed into some other kind of
energy-momentum and absorbed by the external field or dissipated. Since
the two components $F$ and $*F$ are equal in rights it is naturally to expect
that the corresponding 2 terms, defining the exchange in a unit 4-volume,
will depend on $F$ and $*F$ similarly. The above expression for
$\nabla_\nu Q_\mu^\nu$ gives the two 1-forms
\[
F_{\mu\nu}(\delta F)^\nu dx^\mu,\quad  (*F)_{\mu\nu}(\delta* F)^\nu dx^\mu
\]
as natural candidates for this purpose. As for the third channel, it takes
into account a possible influence of the external field on the intra-field
exchange between $F$ and $*F$, which occurs without absorbing of
energy-momentum by the external field. The natural candidate, describing this
exchange is, obviously, the expression
\[
F_{\mu\nu}(\delta *F)^\nu dx^\mu + (*F)_{\mu\nu}(\delta F)^\nu dx^\mu.
\]
It is important to note, that these three channels are independent in the
sense, that any of them may occur without taking care if the other two work
or don't work. A natural model for such a situation is a 3-dimensional vector
space $K$, where the three dimensions correspond to the three exchange
channels. The non-linear exchange law requires some $K$-valued non-linear
map. Since our fields take values in ${\cal R}^2$ this 3-dimensional space
must be constructed from ${\cal R}^2$ in a natural way. Having in view the
bilinear character of $\nabla_\nu Q^\nu_\mu$ it seems naturally to look for
some bilinear construction with the properties desired. These remarks suggest
to choose for $K$ the {\it symmetrized tensor product}
$Sym({\cal R}^2\otimes {\cal R}^2)\equiv {\cal R}^2\vee {\cal R}^2$.
So, from the point of view of the $EM$-field, the energy-momentum exchange
term should be written in the following way:
\begin{equation}
*\vee(\delta \Omega,*\Omega).
\end{equation}
In fact, in the corresponding basis $(e_1,e_2)$ we obtain
\[
*\vee(\delta \Omega,*\Omega)=
*\vee (\delta F \otimes e_1 +\delta *F \otimes e_2, *F\otimes e_1+**F\otimes
e_2)=
\]
\[
=*(\delta F\wedge *F)\otimes e_1\vee e_1
+*(\delta *F\wedge **F)\otimes e_2\vee e_2
+ *(-\delta F\wedge F + \delta*F\wedge *F)\otimes e_1\vee e_2
\]
\[
=F_{\mu\nu}(\delta F)^\nu dx^\mu \otimes e_1\vee e_1 +
(*F_{\mu\nu})(\delta *F)^\nu dx^\mu \otimes e_2\vee e_2 +
\]
\[
+\left[F_{\mu\nu}(\delta *F)^\nu dx^\mu +
(*F_{\mu\nu})\delta F^\nu dx^\mu\right]\otimes e_1\vee e_2.
\]
This expression determines how much of the $EM$-field
energy-momentum may be carried irreversibly over to the external field (the
first and the second terms) and how much may be redistributed between
$F$ and $*F$ by virtue of the external field influence in a unit 4-volume.

Now, this same quantity of energy-momentum has to be expressed by new terms,
in which the external field "agents" should participate. Let's denote  by
$\Phi$ the first agent, interacting with $\pi_1\Omega$, and by $\Psi$ the
second agent, interacting with $\pi_2\Omega$. Since the corresponding two
exchanges are independent, we may write the exchange term in the following
way:
\begin{equation}
\vee(\Phi,*\pi_1\Omega)\ +\ \vee(\Psi,*\pi_2\Omega).
\end{equation}
According to the local energy-momentum conservation law these two
quantities have to be equal, so we obtain
\begin{equation}
\vee(\delta \Omega,*\Omega)=
\vee(\Phi,*\pi_1\Omega)\ +\ \vee(\Psi,*\pi_2\Omega).
\end{equation}

This is the basic relation in EED. It contains the basic differential
equations for the $EM$-field components and requires additional equations,
specifying the properties of the external field, i.e. the algebraic and
differential properties of $\Phi$ and $\Psi$. The physical sense of this
equation is quite clear: local balance of energy-momentum.

The coordinate free written relationship (12) is equivalent to the following
relations: $(\Phi=\alpha^1\otimes e_1+\alpha^2\otimes e_2,
\Psi=\alpha^3\otimes e_3+\alpha^4\otimes e_4)$
\begin{eqnarray}
&\delta F\wedge *F=
\alpha^1\wedge*F,\ \delta *F\wedge **F=\alpha^4\wedge**F,\\ \nonumber
&\delta F\wedge **F+\delta *F\wedge *F=\alpha^3\wedge**F +\alpha^2\wedge *F,
\end{eqnarray}
or, in components (the 1-forms $\alpha^i$ will be called {\it currents} also)
\begin{eqnarray}
&F_{\mu\nu}(\delta F)^\nu=F_{\mu\nu}(\alpha
^1)^\nu,\ (*F)_{\mu\nu}(\delta*F)^\nu=(*F)_{\mu\nu}(\alpha ^4) ^\nu,\\
\nonumber
&F_{\mu\nu}(\delta*F)^\nu+(*F)_{\mu\nu}(\delta
F)^\nu=(*F)_{\mu\nu}(\alpha^3) ^\nu + F_{\mu\nu}(\alpha^2) ^\nu.
\end{eqnarray}
Moving everything on the left, we get
\[
(\delta F-\alpha^1)\wedge *F=0,\ (\delta *F-\alpha^4)\wedge **F=0,\
\]
\[
(\delta F-\alpha^3)\wedge **F+(\delta *F-\alpha^2)\wedge *F=0,
\]
or in components
\[
F_{\mu\nu}(\delta F-\alpha^1)^\nu=0,\
(*F)_{\mu\nu}(\delta*F-\alpha^4)^\nu=0,\
\]
\[
F_{\mu\nu}(\delta*F-\alpha^2)^\nu+(*F)_{\mu\nu}(\delta F-\alpha^3)^\nu=0.
\]
Summing up the two equations
\[
F_{\mu\nu}(\delta F)^\nu=F_{\mu\nu}(\alpha ^1)^\nu,
\ (*F)_{\mu\nu}(\delta*F)^\nu=(*F)_{\mu\nu}(\alpha ^4) ^\nu
\]
we obtain
\[
F_{\mu\nu}(\delta F)^\nu+(*F)_{\mu\nu}(\delta*F)^\nu=\nabla_\nu Q_\mu^\nu=
F_{\mu\nu}(\alpha ^1)^\nu+(*F)_{\mu\nu}(\alpha ^4) ^\nu.
\]
This relation shows that the sum
\[
F_{\mu\nu}(\alpha ^1)^\nu+(*F)_{\mu\nu}(\alpha ^4) ^\nu
\]
is a divergence of a 2-tensor, which we denote by  $-P_\mu^\nu$. In this way
we obtain the local conservation law
\begin{equation}
\nabla_\nu (Q_\mu^\nu +P_\mu^\nu)=0.                    
\end{equation}
Thus, we get the possibility to introduce the full energy-momentum tensor
\[
T_\mu^\nu = Q_\mu^\nu +P_\mu^\nu,
\]
where $P_\mu^\nu$ is interpreted as {\it interaction energy-momentum tensor}.
Clearly, $P_\mu^\nu $ can not be determined uniquely in this way.

So, according to (14), for the 22 functions  $F_{\mu\nu}, (\alpha^i)_\mu$
we have 12 equations, and these 12 equations are differential with respect to
$F_{\mu\nu}$ and algebraic with respect to $(\alpha^i)_\mu$. Our purpose now
is to try to write down differential equations for the components of the 4
currents $\alpha^i$. The leading idea in pursuing this goal will be to
establish a correspondence between the physical concept of {\it
non-dissipation} and the mathematical concept of {\it integrability of Pfaff
system}. The suggestion to look for such a correspondence comes from the
following considerations.

Recall from the theory of the ordinary differential equations (or vector
fields), that every solution of a system of ordinary differential equations
(ODE) defines a local (with respect to the parameter on the trajectory) group
of transformations, frequently called {\it local flow}. This means, in
particular, that the motion along the trajectory is admissible in the two
directions: we have a {\it reversible} phenomenon, which has the physical
interpretation of {\it lack of losses} (energy-momentum losses are meant).
Assuming this system of ODE describes {\it fully} the process of motion of a
small piece of matter (particle), we assume at the same time, that {\it all}
energy-momentum exchanges between the particle and the outer field are taken
into account, i.e. we have assumed that {\it there is no dissipation}. In
other words, {\it the physical assumption for the lack of dissipation is
mathematically expressed by the existence of a solution - local flow, having
definite group properties}. The existence of such a local flow is guaranteed
by the corresponding theorem for existence and uniqueness of a solution at
given initial conditions. This correspondence between the mathematical fact
{\it integrability} and the physical fact {\it lack of dissipation} in the
simple case "motion of a particle" , we want to generalize in an appropriate
way, having in view possible applications in more complicated physical
systems, in particular, the physical situation we are going to describe:
interaction of the field $\Omega $ with some outer field, represented in
the exchange process by the four 1-forms $\alpha ^i$. This will allow to
write down equations for $\alpha^i$ in a direct way. Of course, in the real
world there is always dissipation, and following this idea we are going to
take into account its neglecting as conditions (i.e. equations) on the
currents $\alpha^i$. As it is well known,
the mathematicians have made serious steps towards study and formulation of
criteria for integrability of partial differential equations, so it looks
unreasonable to close eyes before the available and represented in
appropriate form mathematical results.

\vskip 0.5cm
\noindent{\it Remark}. It is interesting, and may be suggesting, to note the
following . In physics we have two {\it universal} things: {\it dissipation}
and {\it gravitation}. We are going now to establish a correspondence between
the physical notion of {\it dissipation} and the mathematical concept of {\it
non-integrability}. As we know, the mathematical  non-integrability is
measured by the concept of {\it curvature}. General theory of Relativity
describes gravitation by means of Riemannian curvature. The circle will
close if we connect the universal property of any real physical  process
to dissipate energy-momentum with the only known so far universal
interaction in nature, the gravitation.

\vskip 0.5cm
First we note that our base manifold, where all fields and operations are
defined, is the simple 4-dimensional Minkowski space. According to our
equations (12) the medium reacts to the field $\Omega $ by means of the two
${\cal R}^2$-valued 1-forms $\Phi=\alpha^1\otimes e_1 +\alpha^2\otimes e_2$
and $\Psi=\alpha^3\otimes e_1 +\alpha^4\otimes e_2$. So, we obtain four
${\cal R}$-valued 1-forms $\alpha^1, \alpha^2, \alpha^3, \alpha^4$. Because
of the 4-dimensions of Minkowski space it
is easily seen that only 1-dimensional and 2-dimensional Pfaff systems may be
of interest from the Frobenius integrability point of view. All Pfaff systems
of higher dimension are trivially integrable.

The integrability equations for 1-dimensional Pfaff systems are
\begin{equation}
{\bf d}\alpha^i\wedge \alpha^i=0,\ i=1,2,3,4.                  
\end{equation}
Every of the 4 equations (16) is equivalent to 4 scalar nonlinear equations
for the components of the corresponding 1-form. We note, that the solutions
of (16), as well as the solutions of the general integrability equations
for a $p$-dimensional Pfaff system, are determined up to a scalar multiplier,
i.e. if $\alpha^i$ define a solution, then $f_i.\alpha^i$
(no summation over $i$),
where $f_i$ are smooth functions, define also a solution.

In case of 2-dimensional Pfaff systems $(\alpha^i,\alpha^j)$, defined by four
1-forms, their maximal number is 4.3=12. The Frobenius equations read
\begin{equation}
{\bf d}\alpha^i\wedge \alpha^i\wedge \alpha^j=0,\ i\neq j.     
\end{equation}
We have here 12 nonlinear equations for the all 16 components of $\alpha^i$,
\linebreak $i=1,2,3,4$.
Clearly, these equations (17) are substantial only if the corresponding
$\alpha^i$, the exteriour differential ${\bf d}\alpha^i$ of which
participates in (17), does not satisfy (16) or is not zero.

Our assumption now reads:
\vskip 0.05cm
{\it Every 2-dimensional Pfaff system, defined by
the four 1-forms $\alpha^i$ is \linebreak
\indent completely integrable}.
\vskip 0.05cm
As we mentioned above, physically this assumption means that we neglect the
dissipation of energy-momentum. Note also, that 1-dimensional nonintegrable
Pfaff systems are admissible, which physically means, that if there is some
1-dimensional nonintegrable Pfaff system among the $\alpha^i$, e.g.
$\alpha^1$, then the corresponding dissipation of energy-momentum does not
flow out of the physical system and it is utilized by the exchange processes,
described by the rest currents $\alpha^2,\alpha^3,\alpha^4$.

Finally we note, that (14) and (16) give in general 24 equations for the 22
functions $F_{\mu\nu}, (\alpha)^i_\mu$, which seem to be enough to obtain the
dynamics of the system at given initial condition. As for the Maxwell's
theory, it corresponds to $\pi_2\delta\Omega=0$ and
$\pi_1\delta\Omega=4\pi (J_{free}+J_{bound})\otimes e_1$,
i.e. $\alpha^2=\alpha^4=0$ and
$\alpha^1=\alpha^3=4\pi(J_{free}+J_{bound})$.

\vskip 0.5cm
In conclusion we note, that our notion of $EM$-field requires a simultaneous
consideration of a number of soliton-like solutions and, in particular, a
possible interaction (interference) among them. This problem does not fit
to the goal we set in this paper, so it will be considered somewhere else.

\newpage
{\bf \large References}
\vskip 0.5cm
[1]. {\bf J.D.Jackson}, {\it CLASSICAL ELECTRODYNAMICS}, John Wiley and Sons,
Inc., New York-London, 1962., also its new addition, and all cited therein
monographs on CED.

[2]. {\bf J.Plebansky}, {\it LECTURES ON NONLINEAR ELECTRODYNA\-MICS},
Nordita, 1970.

[3]. {\bf M.Born, L.Infeld}, Proc.Roy.Soc., A 144 (425), 1934

[4]. {\bf A.Lees}, Phyl.Mag., 28 (385), 1939

[5]. {\bf N.Rosen}, Phys.Rev., 55 (94), 1939

[6]. {\bf P.Dirac}, Proc.Roy.Soc., A268 (57), 1962

[7]. {\bf H.Schiff}, Proc.Roy.Soc., A269 (277), 1962

[8]. {\bf H.Schiff}, Phys.Rev., 84 (1), 1951

[9]. {\bf D.Finkelstein, C.Misner}, Ann.Phys., 6 (230), 1959

[10]. {\bf D.Finkelstein}, Journ.Math.Phys., 7 (1218), 1966

[11]. {\bf J.P.Vigier}, Found. of Physics, vol.21 (1991), 125.

[12]. {\bf T.Waite}, Annales de la Fondation Louis de Broglie, 20 No.4, 1995.

[13]. {\bf H.Cartan}, {\it CALCUL DIFFERENTIEL. FORMES DIFFERENTI\-ELLES},
Herman, Paris, 1967

[14]. {\bf C.Godbillon}, {\it GEOMETRY DIFFERENTIELLE ET MECANIQUE
\\ANALYTIQUE}, Herman, Paris 1969

[15]. {\bf S.Donev}, {\it A particular non-linear generalization of Maxwell
equations admitting spatially localized wave solutions}, Compt.Rend.Bulg.
Acad.\newline Sci., vol.34, No.4, 1986

[16]. {\bf S.Donev}, {\it A covariant generalization of sine-Gordon equation
on \newline Minkowski space-time}, Bulg.Journ.Phys., vol.13, (295), 1986

[17]. {\bf S.Donev}, {\it Autoclosed differential forms and (3+1)-solitary
waves},\newline Bulg.Journ.Phys., vol.15, (419), 1988

[18]. {\bf S.Donev}, {\it On the description of single massless quantum
objects}, Helvetica Physica Acta, vol.65, (910), 1992

[19]. {\bf S.Donev, M.Tashkova}, {\it Energy-momentum directed
nonlinearization of Maxwell's pure field equations}, Proc.R.Soc.of Lond., A
443, (301), 1993

[20]. {\bf S.Donev, M.Tashkova}, {\it Energy-momentum directed
nonlinearization of Maxwell's equations in the case of a continuous medium},
Proc.R.Soc. of Lond., A 450, (281), 1995

\end{document}